\documentclass[aps,pra,onecolumn,superscriptaddress,10pt]{revtex4}

\usepackage{latexsym}
\usepackage{amsfonts}
\usepackage{amsmath}
\usepackage{dcolumn}
\usepackage{url}

\usepackage{setspace}
\usepackage{graphicx}

\DeclareMathAlphabet{\mathpzc}{OT1}{pzc}{m}{it}
\newcommand\Real{\mbox{Re}} 
\newcommand\Imag{\mbox{Im}} 

\newsavebox{\astrutbox}
\sbox{\astrutbox}{\rule[-5pt]{0pt}{20pt}}

\newcommand{\rmd}{{\mathrm{d}}}
\newcommand{\rme}{{\mathrm{e}}}
\newcommand{\rmi}{{\mathrm{i}}}

\newcommand{\cC}{{\mathcal C}}
\newcommand{\cD}{{\mathcal D}}
\newcommand{\cF}{{\mathcal F}}
\newcommand{\cG}{{\mathcal G}}

\newcommand{\cZ}{{\mathcal Z}}

\newcommand{\cM}{{\cal M}}
\newcommand{\summ}{\displaystyle\sum}

\begin{document}

\def\Brasilia{Instituto de F\'{\i}sica, Universidade de Bras\'{\i}lia, \\
	CP: 04455, 70919-970 Bras\'{\i}lia, DF, Brasil}
\def\Marseille{Aix-Marseille Universit\'{e} and CNRS, UMR 7345 PIIM, \\
   case 322, campus Saint-J\'er\^ome, FR-13013 Marseille, France}

\title{Phase mixing importance for both Landau instability and damping}

\author{ D~D~A Santos}
\affiliation{\Brasilia}
\email{daniel_dourado@yahoo.com.br}

\author{Yves Elskensl}
\affiliation{\Marseille}
\email{yves.elskens@univ-amu.fr}





\begin{abstract}
We discuss the self-consistent dynamics of plasmas by means of hamiltonian formalism 
for a system of $N$ near-resonant electrons  
interacting with a single Langmuir wave. 
The connection with the Vlasov description is revisited 
through the numerical calculation of the van Kampen-like eigenfrequencies 
of the linearized dynamics for many degrees of freedom. 
Both the exponential-like growth as well as damping of the Langmuir wave 
are shown to emerge from a phase mixing effect among beam modes,  
revealing unexpected similarities between the stable and unstable regimes.


\medskip \textbf{Keywords} : phase mixing, Landau damping/growth, monokinetic beams, eigenfrequencies continuum, van Kampen modes 

{\textbf{PACS numbers}} :     \newline  
  52.20.-j	 Elementary processes in plasmas \newline
  45.50.-j	 Dynamics and kinematics of a particle and a system of particles \newline
  05.20.Dd	  Kinetic theory \newline 
  52.35.Fp	  Electrostatic waves and oscillations  \newline

\end{abstract}

\maketitle





\section{Introduction}
\label{sec:Intro}

Collisionless damping of electrostatic waves consists one of the most fundamental phenomena in plasma physics
and a starting point for the studies on the wave-particle interaction.
From its first report by \cite{Landau1946}
until its experimental verification by \citet{Malmberg1964},
this phenomenum remained under great caution by the plasma community,
mainly due to the mathematical aspects (specifically the use of complex integral and
the large time limit in the Laplace transform) of its derivation and its paradoxical nature.

A very physical and intuitive approach was proposed by \citet{Dawson1961}, 
who explained this effect as resulting from a near-resonance mechanism of energy
(and momentum) transfer between wave and particles.
Dawson also pointed out a time threshold for a breakdown (due to trapping effects) of the Landau's linear analyses.
Later on, \cite{ONeil1965} extendend the Landau damping rate $\gamma_\mathrm{L}$ for arbitrary times and
found that the nonlinearities may lead the wave to a time-asymptotic behaviour with a constant nonzero amplitude.
O'Neil's picture introduced in the sixties, gave rise to what we call today ``nonlinear Landau damping''
and propelled a reconsideration of the phenomenum on experimental \citep{Malmberg1967,Franklin1972}, numerical \citep{Brodin1997,Manfredi1997} and theoretical \citep{MouVil2010,MouVil2011} stages.
In the last decades, the existence of a critical initial amplitude that distinguish wether the asymptotic field Landau damps to zero or evolves to a steady value has been proved, by solving the Vlasov-Poisson system,
by \cite{Brunetti2000} and \cite{LancelDorning2003} and
in a self-consistent (hamiltonian) way by \cite{FirEls2000} and \cite{FirEls2001}.

Another prominent interpretation for linear Landau damping (beside Dawson's mechanical picture) 
explains it through the phase mixing of generalized (singular) eigenmodes of the corresponding linear dynamical system. 
This formulation, resting on a strictly linear theory and originally proposed by \citet{vanKampen1955}, 
shows that for a given perturbation with wavenumber $k$ there is a continuous spectrum of real frequencies allowed for the plasma oscillations. 
In this context, Landau damping emerges from a special superposition of stationary (distribution-like) solutions 
that now go under the name of van Kampen modes. 
Though such superpositions account for the exponential decay of initial perturbations at Landau's rate,
special initial data can also be constructed which cause a decay at \emph{slower} rates \citep{Belmont2011}.

Nowadays, it is well known that other kinds of systems also admit normal modes 
analogous to the van Kampen modes for plasmas 
\citep[see e.g.][]{Pen1994,Vandervoort2002,Vekstein1998}. 
Indeed, similar phenomena are found in a wide variety of systems 
such as e.g.\ dusty plasmas \citep{Bliokh1995}, galaxies \citep{Kandrup1998},
two-dimensional fluid flows \citep{CastilloFirpo2002} 
and liquids containing gas bubbles \citep{Smereka2002}. 
Despite its being an old problem, Landau damping remains intensely investigated 
experimentally \citep{DovEscMac2005} and theoretically 
\citep[see e.g.][]{Dougherty1998, Ryutov1999, del-Castillo-Negrete-Houches2002, Elskens2005Fields, MouVil2010, MouVil2011, Bratanov_etal2013, Benisti2016} 
with new aspects and viewpoints still being explored.

In contrast with the difficult reception of Landau damping by the physics community, 
the weak warm beam instability, captured in the same formula (with an opposite sign reflecting its physics), 
was immediately accepted. We show in this paper that its status is somewhat more subtle, 
and that this Landau instability has more to share with damping than might be thought initially. 
Indeed, the aim of this paper is to provide further theoretical and numerical insight into 
how both Landau effects emerge from the phase mixing among the van Kampen-like modes 
in the hamiltonian approach. 

The key element in plasmas is their being an $N$-body system.\cite{note1} 
Therefore, a microscopic description involving actual particles, 
rather than a continuum represented by a smooth distribution function in $({\mathbf{r}}, {\mathbf{v}})$ space,
is their physical fundamental model \citep{EEbook2003, EED_Vlasovia}. 
As deriving a kinetic model for singular interactions, with the Coulomb divergence at short range, 
remains a challenge \citep{Kie14, Chaffi14, BrenigChaffi16}, 
understanding basic plasma phenomena from the $N$-body picture is important. 
Moreover, the $N$-body approach yields its own benefits, such as showing unexpected connections between 
Landau damping and Debye screening \citep{EDE_PPCF2016}. 

The formalism adopted here is closely related to the fluid model used by \citet{Dawson1960}, 
where electrons are distributed into (almost) monokinetic beams, 
and the central issue consists in analysing normal modes that fully describe the evolution of the system. 
This approach is mathematically elementary, involving no partial differential equations, nor functional analysis,
and not even analytic continuation or $\rmi \varepsilon$ prescription enforcing a causality argument. 
Its concepts belong in basic classical mechanics. 

In section~\ref{sec:Formalism}, 
we revisit the hamiltonian model for plasmas and waves \citep{EscZekEls1996, EEbook2003} 
and discuss connections with the van Kampen--Case approach \citep{vanKampen1955, Case1959} 
when the continuous limit is taken. 
In section~\ref{sec:Results}, 
we obtain the spectrum of the discrete analogue to the van Kampen frequencies 
by solving accurately the dispersion relation for systems composed of up to $2000$ beams. 
We also investigate differences between the damping and growth regimes 
and the consistency between the discrete and the continuous systems 
by monitoring the evolution of the wave intensity.

Our main results show that not only the damping but also the Landau instability
emerge as consequence of a phase mixing mechanism among the eigenmodes of the linear system.
We highlight that while in the stable case
the phase mixing to generate Landau damping involves all van Kampen-like eigenmodes, 
in the unstable case the pure Landau growth results from a destructive interference effect leaving 
a single eigenmode having a dominant (exclusive, in the continuum limit) contribution to the wave amplitude.
This behaviour, observed for dense spectrum system,
is described through equation \eqref{eq:Z_decomposed} and its asymptotic form given in \eqref{eq:Z-1}
and illustrated in Figures~\ref{fig:cancellation} and \ref{fig:evolution}. 


The outcomes reported in this paper for systems with many degrees of freedom
were made possible only by the development of a new technique to compute complex roots. 
This new root finding method is further detailed in appendix~\ref{sec:AppA}.


\section{Formalism of monokinetic beams}
\label{sec:Formalism}

Our model system is composed of $N$ charged resonant particles 
interacting with a single electrostatic Langmuir wave with natural frequency $\omega_0$, in one space dimension 
\citep{Onishchenko70, ONeil71}. 
The evolution of this system is generated by the self-consistent hamiltonian \citep{Mynick78, Tennyson94, EscZekEls1996, EEbook2003,Escande2010Houches} 
\begin{equation} \label{eq:Hsc}
	H_{\mathrm{sc}} 
	= 
	\displaystyle\sum_{l = 1}^{N} \frac{p^2_l}{2} 
	+ \displaystyle \omega_{0} \frac{X^2 + Y^2}{2} 
	+ \varepsilon k_{\mathrm{w}}^{-1} \displaystyle\sum_{l = 1}^{N} 
	                   \left(Y \sin{k_{\mathrm{w}} x_l} - X \cos{k_{\mathrm{w}} x_l}\right),
\end{equation}
\noindent where $X$ and $Y$ correspond to the Cartesian components of the complex wave amplitude $Z = X + \rmi Y$, 
also expressed as $Z = \sqrt {2I} \ \rme^{-\rmi \theta}$ in terms of the phase $\theta$ and intensity $I$. 
We assume that particles have periodic boundary conditions on the interval of length $L$, 
and the wave number is $k_{\mathrm{w}} = 2\pi j /L$ with some integer  $j$.
Parameter $\varepsilon$ is the wave-particle coupling constant 
(which may be determined from further considerations on the underlying plasma). 

The first and second terms in \eqref{eq:Hsc} represent the kinetic energy of the resonant particles 
(generating ballistic motion) 
and the energy of the free wave (related to the vibratory motion of the non-resonant bulk particles in the plasma), respectively. 
The last one (also responsible for the nonlinear nature of the dynamics) 
corresponds to the interaction energy between the wave and resonant particles.

The equations of motion are directly derived from the hamiltonian \eqref{eq:Hsc}, 
\begin{eqnarray}
	\label{eq:xdot} \dot{x}_l &=& p_l~, \hskip31mm 1 \leq l \leq N , \\
	\label{eq:pdot} \dot{p}_l &=& \varepsilon~\Imag\left(Z \ \rme^{\rmi k_{\mathrm{w}} x_l} \right)~, \hskip1cm 1 \leq l \leq N , \\
 	\label{eq:Zdot} \dot{Z} 
	&=& -\rmi \omega_{0} Z + 
	        \rmi \varepsilon k_{\mathrm{w}}^{-1} \displaystyle\sum_{l^{\prime} = 1}^{N} \ \rme^{-\rmi k_{\mathrm{w}} x_{l^{\prime}}}~. 
\end{eqnarray}
This dynamical system admits as equilibrium states the configurations where the electrostatic field has zero amplitude 
and the particles are distributed  
in monokinetic beams (as in Dawson's beams model), labeled $1 \leq s \leq b$,  
characterized by velocities $v_s$ and number of particles $N_s$. 
Assuming
\begin{equation}
  x^{(0)}_{ns}(t) = v_s t + n L/N_s + \phi_s  \label{eq:free-motion}
\end{equation} 
as the position of the $n$-th particle of beam $s$ with number of particles per beam satisfying the condition $N_s > j$, 
the sum in equation \eqref{eq:Zdot} vanishes at any instant in time and, consequently, 
$(x_l(t) = x^{(0)}_{ns}(t), p_l(t) = v_s, Z(t) = 0)$ corresponds to an exact, equilibrium solution with vanishing wave. 
No similar solution exists with nonzero $Z$ \citep{Elskens2001}.

The dynamics preserves total energy $H$ and total momentum $P = k_{\mathrm{w}} I + \sum_l p_l$. 
The reversibility of hamiltonian dynamics is expressed 
by the invariance of system \eqref{eq:xdot}-\eqref{eq:pdot}-\eqref{eq:Zdot} 
under the time-reversal map $(x', p', X', Y', t', H', k_{\mathrm{w}}') = (x, -p, -X, Y, -t, H, -k_{\mathrm{w}})$. 
Here, the reversal of $k_{\mathrm{w}}$ 
accounts for the reversal of total momentum $P$ under this map
as well as invariance of total energy $H$. 

Considering the system slightly displaced from this equilibrium configuration, 
the evolution of the wave amplitude and the perturbations in particle orbits 
obey the linearized equations of motion
\begin{eqnarray}
\label{eq:dx} \delta \dot{x}_{ns} &=& \delta p_{ns}~, \\
\label{eq:dp} \delta \dot{p}_{ns} &=& -\varepsilon~\Imag\left( Z \ \rme^{-\rmi k_{\mathrm{w}} x^{(0)}_{ns}} \right)~, \\
\label{eq:dZ} \dot{Z} &=& -\rmi \omega_0 Z + \varepsilon \displaystyle\sum_{s = 1}^{b} \displaystyle\sum_{n = 1}^{N_s} \delta x_{ns} \ \rme^{-\rmi k_{\mathrm{w}} x^{(0)}_{ns}}~, 
\end{eqnarray}
\noindent where $b$ is the number of beams ($\sum_{s = 1}^{b} N_s = N$ ), $\delta x_{ns} = x_{ns} - x^{(0)}_{ns}$ and $\delta p_{ns} = p_{ns} - v_s$.

The perturbations in positions and momenta can be expressed in Fourier series with coefficients $C$ and $A$ \citep{EscZekEls1996},
\begin{eqnarray}
\label{eq:dx-Fourier2}
  \delta x_{ns}(t) 
  &=& 
  2 \, \Imag\left(C_{s}(t) \ \rme^{\rmi k_{\mathrm{w}} x^{(0)}_{ns}(t)}\right) - \rmi \displaystyle\sum_{m \in \mu^{\prime}_s} C_{ms}(t) \ \rme^{\rmi k_m x^{(0)}_{ns}(t)}~, \\
\label{eq:dp-Fourier2} 
  \delta p_{ns}(t) 
  &=& 
  2 \, \Real\left(A_{s}(t) \ \rme^{\rmi k_{\mathrm{w}} x^{(0)}_{ns}(t)}\right) + \displaystyle\sum_{m \in \mu^{\prime}_s} A_{ms}(t) \ \rme^{\rmi k_m x^{(0)}_{ns}(t)}~, 
\end{eqnarray}
\noindent where the set $\mu^{\prime}_s$ is defined as
\begin{equation} \label{eq:mus}
  \mu^{\prime}_s = \left\{
     \begin{array}{@{}l@{\thinspace}l}
		\left\{m \in \mathbb{Z} : |m| \leq (N_s - 1)/2 \ {\mathrm{and}} \ m \neq \pm j \right\}, & \mbox{if $N_s$ is odd,}\\
		\left\{m \in \mathbb{Z} : 1 - N_s/2 \leq m \leq N_s/2 \ {\mathrm{and}} \ m \neq \pm j \right\}, & \mbox{if $N_s$ is even.}\   
     \end{array}
   \right.
\end{equation}
The first term in equations \eqref{eq:dx-Fourier2} and \eqref{eq:dp-Fourier2}, which appears outside the sum, 
corresponds to what we call the wavelike part of the solution. 
We omitted subscript $j$ in its coefficients since it is the only component of the wavelike kind. 
The second term, involving the sums over $\mu^{\prime}_s$, 
corresponds to the ballistic part whose evolution is readily expressed in terms of initial conditions, 
\begin{eqnarray}
\label{eq:C-ball} C_{ms}(t) &=& (C_{ms}(0) + \rmi A_{ms}(0) t) \ \rme^{-\rmi k_m v_s t} \\
\label{eq:A-ball} A_{ms}(t) &=& A_{ms}(0) \ \rme^{-\rmi k_m v_s t}~,
\end{eqnarray}
because those terms do not couple with $Z$ as $\sum_n \rme^{-i k_m x_{ns}^{(0)}} = 0$ for each $m, s$.
To the contrary, due to its dependence on the wave amplitude $Z$, 
the wavelike component of the Fourier coefficients has a more intricate form 
that requires calculating the eigenvalues of the linearized system and expanding initial data in terms of the corresponding eigenvectors. 


\subsection{Wave-like solution and dispersion relation}
\label{sec:w-l sol and disp rel}

The ``lattice'' Fourier coefficients of the wavelike part satisfy the system of differential equations
\begin{eqnarray}
\label{eq:C-wave} \dot{C}_{s} &=& -\rmi k_{\mathrm{w}} v_s C_{s} + \rmi A_{s}~,  \qquad 1 \leq s \leq b,\\
\label{eq:A-wave} \dot{A}_{s} &=& -\rmi k_{\mathrm{w}} v_s A_{s} + \rmi \frac{\varepsilon}{2} Z~,  \qquad 1 \leq s \leq b,\\
\label{eq:Z-wave} \dot{Z} &=& \rmi \omega_0 Z - \rmi\varepsilon\summ_{s = 1}^{b} C_{s} N_s~.
\end{eqnarray}

\begin{figure}
	\centering
	\begin{minipage}[t]{.49\linewidth}
		\centering
		\includegraphics[width = \linewidth]{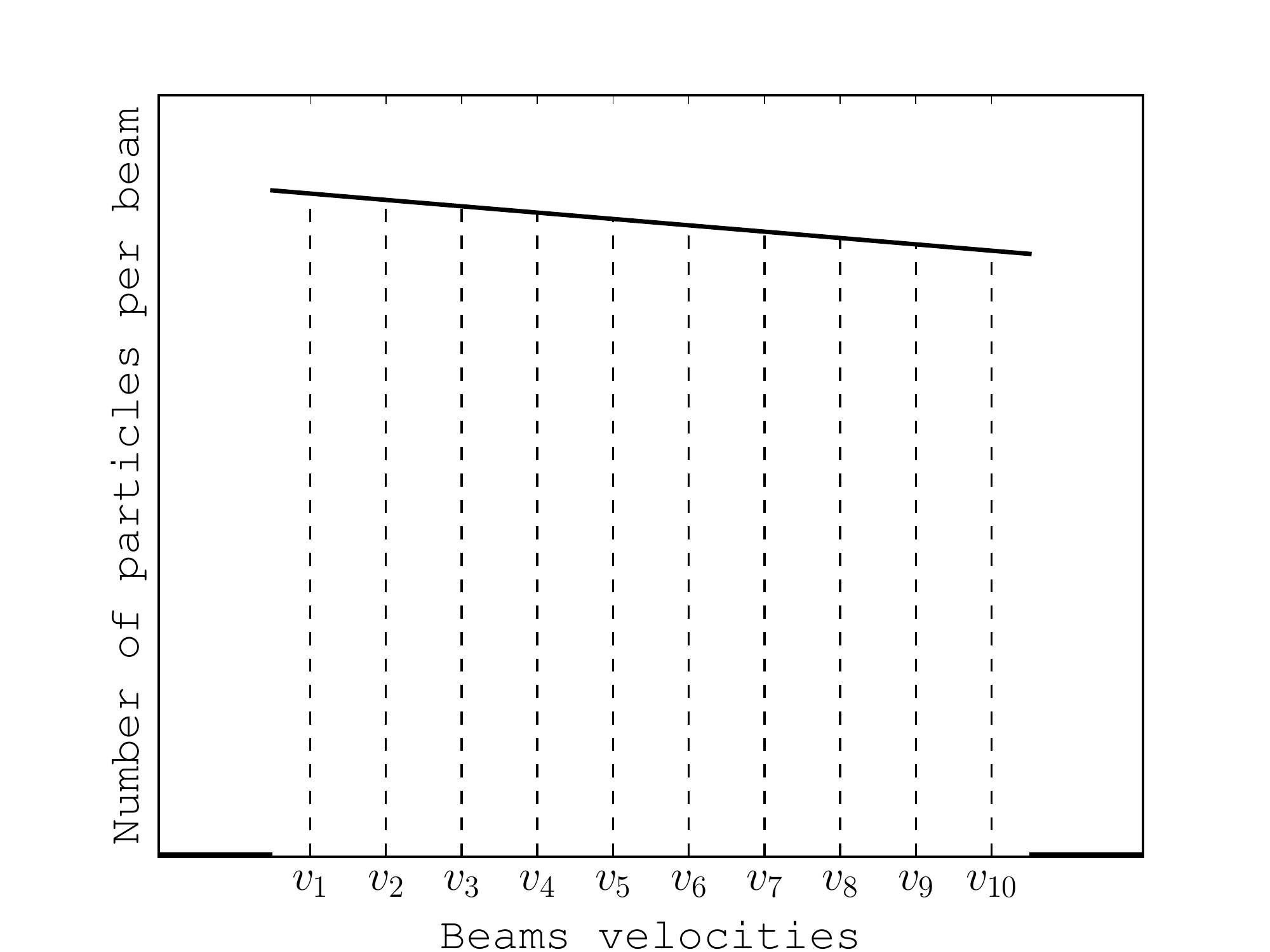}
		\label{fig:distribution-velocities}
	\end{minipage}%
	\begin{minipage}[t]{.49\linewidth}
		\centering
		\includegraphics[width = \linewidth]{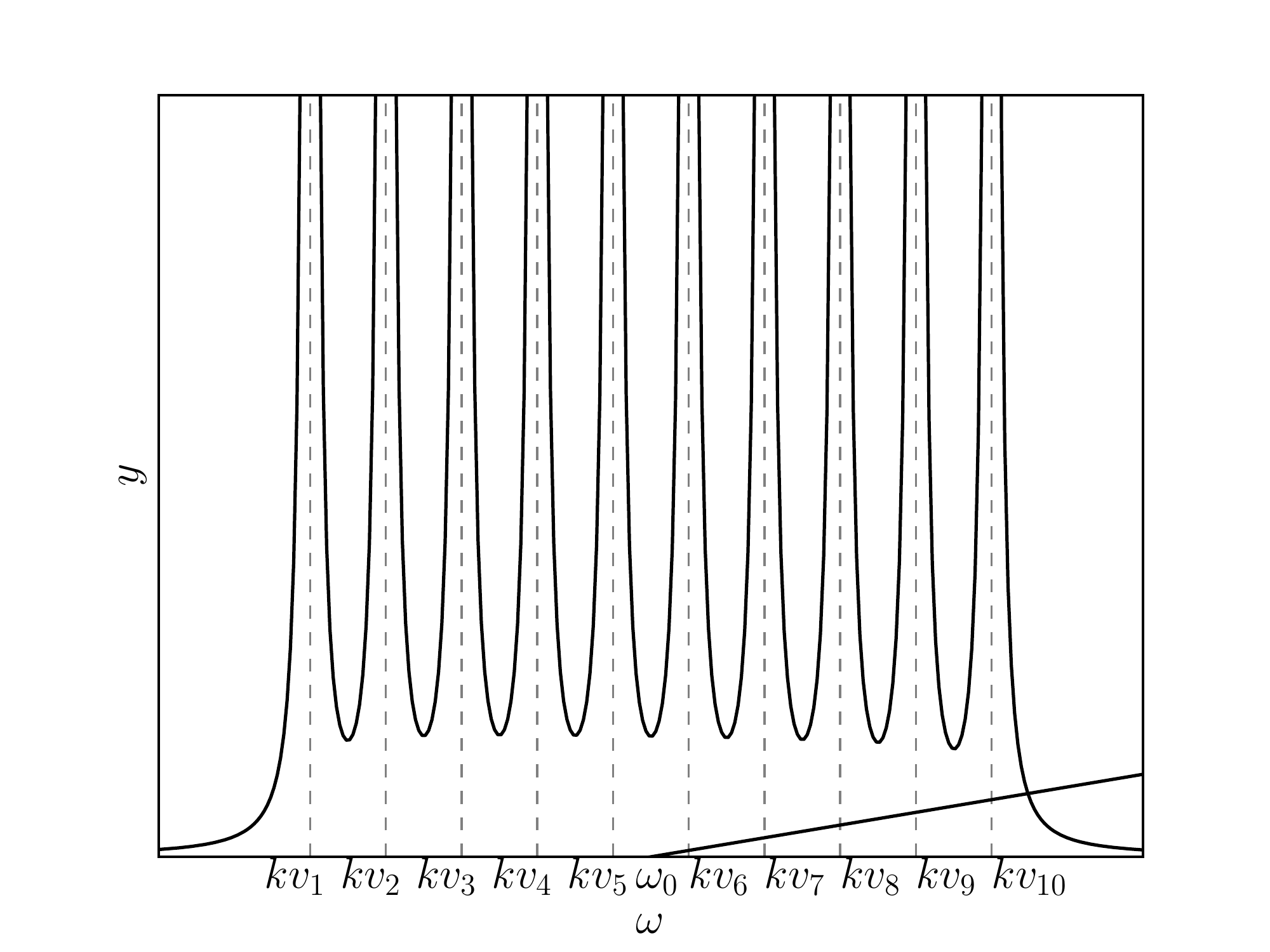} 
		\label{fig:dispersion-relation}	
	\end{minipage}
	\caption{Illustration of a possible configuration for a system composed of one wave
	with natural frequency $\omega_0$ and ten monokinetic beams.
	The beams velocities $v_s = v_1 + (s - 1) \Delta p$ are set around the wave phase velocity $\omega_0/k$.
	(left) Distribution of particles velocities with a constant slope, given according to
	$f(v_s) = f(v_1) + (s - 1) \Delta p f'$ (yielding $N_s = N \Delta p f(v_s)$ as the number of particles)
	in $I_v = [v_1 - \Delta p/2,v_b + \Delta p/2]$ and set to zero outside this interval.
	(right) Graphical representation of the dispersion relation \eqref{eq:dispersion-relation} through the line 		$y = \omega - \omega_{0}$ and the many branches curve $y = \chi(\omega) \equiv (\varepsilon^2/2) \sum_s (\omega - k v_s)^{-2} N_s$. 	The intercept point locates the single purely real solution to this equation.}
	\label{fig:f-dispersion-relation}
\end{figure}

Assuming solutions of the form $\sim \rme^{-\rmi \sigma t}$, 
the problem of solving the $2b + 1$ equations of motion reduces to solving the algebraic linear system
\begin{equation} \label{eq:linear-system}
	\sigma \cC = \cM \cdot \cC,
\end{equation}
\noindent where $\cC \in \mathbb{C}^{2b + 1}$ denotes $\cC = [c_{1}, \dots ,c_{b}, a_{1}, \dots, a_{b}, z]^{\top}$,  whose components $c_s$'s and $a_s$'s represent, respectively, the time-independent part of the Fourier coefficients for the positions and momenta. The real matrix $\cM \in \mathbb{R}^{(2b+1)\times(2b+1)}$ reads
\begin{equation}
	\label{eq:Mj}
	\cM = \left( \begin{array}{ccccccc}
    k_{\mathrm{w}} v_1 &        & 0       & -1 &        & 0  & 0       \\
            & \ddots &         &    & \ddots &    & \vdots  \\
     0      &        & k_{\mathrm{w}} v_b & 0  &        & -1 & 0	    \\
     0  &        & 0 & k_{\mathrm{w}} v_1 &        &    0    & -\varepsilon/2 \\
        & \ddots &   &         & \ddots &         &	\vdots					   \\
     0  &        & 0 &  0      &        & k_{\mathrm{w}} v_b & -\varepsilon/2 \\
    \varepsilon N_1 & \dots & \varepsilon N_b & 0 & \dots & 0 & \omega_{0}     
\end{array} \right)~.
\end{equation}
Denoting by $\cC_r = [c_{r1}, \dots ,c_{rb}, a_{r1}, \dots, a_{rb}, z_r]^{\top}$ the $r$-th eigenvector of $\cM$, 
its associated eigenvalue $\sigma_r$ is obtained through the characteristic equation
\begin{equation} 
\label{eq:dispersion-relation}
  \sigma_r 
  = 
  \omega_0 + \frac{\varepsilon^2}{2} \summ_{s = 1}^{b} \frac{N_s}{(\sigma_r - k_{\mathrm{w}} v_s)^2}~.
\end{equation}
For a given wavenumber $k_{\mathrm{w}}$, equation \eqref{eq:dispersion-relation} states the condition 
to be satisfied by the eigenfrequencies $\sigma_r$ for the system \eqref{eq:C-wave}-\eqref{eq:Z-wave} to admit non-trivial solutions.
This equation is thus a dispersion relation for the eigenmodes of the linearized system. 
It can be transformed into a polynomial equation of degree $2b + 1$ with real coefficients, 
and therefore admits generically $2b + 1$ complex roots, with at least one of them purely real 
(see figure~\ref{fig:dispersion-relation}).

The continuous limit corresponds to letting $N \to \infty$ while keeping $\varepsilon^2 N$ fixed.
Particles are described with the distribution function \citep{Firpo1998}
\begin{equation}
  \cF (x,p) = \lim_{N \to \infty}  \frac{1}{N} \sum_l \delta (x - x_l) \delta (p - p_l)
  \label{eq:flim}
\end{equation}
and the wave is rescaled to 
\begin{equation}
  \cZ = \varepsilon Z
  \label{eq:Zlim}
\end{equation}
so that the dynamics \eqref{eq:xdot}-\eqref{eq:pdot} generates the characteristics of the Vlasov equation
\begin{equation}
  \partial_t \cF + p \, \partial_x \cF - \Imag\left( \cZ \ \rme^{-\rmi k_{\mathrm{w}} x} \right) \partial_p \cF 
  = 0
  \label{eq:Vlasov}
\end{equation}
coupled with the wave evolution
\begin{equation}
 	\label{eq:Zdotlim} 
	\dot{\cZ} 
	= - \rmi \omega_{0} \cZ 
	   + \rmi k_{\mathrm{w}}^{-1} \int_0^L \int_{- \infty}^\infty \rme^{-\rmi k_{\mathrm{w}} x} \cF(x,p) \, \rmd p \, \rmd x~. 
\end{equation}
The equilibrium reference state is $(\cZ = 0, \, \cF(x,v) = L^{-1} f(v))$, 
with the velocity distribution function such that 
$N_s = N \int_{v_s - \Delta p/2}^{v_s + \Delta p /2} f(p) \, \rmd p$~:
in the continuous limit, beam velocities are continuously distributed.

The dispersion relation in this limit follows from \eqref{eq:dispersion-relation} 
by replacing the sum over the number of particles with an integral over the interval containing the beams velocities
weighted by the normalized distribution $f(v_s)$. 
Formally, this leads to a singular integral equation if $\sigma$ is real and $f(\sigma / k_{\mathrm{w}})$ does not vanish~; 
with an $\rmi \epsilon$ prescription, \citet{Landau1946} obtains a solution $\sigma_{\mathrm{L}}$ with imaginary part 
\begin{equation} 
\label{eq:Landau-growth-rate}
  \gamma_{\mathrm{L}} 
  = 
  \frac{\pi \varepsilon^2 N}{2 k^2_{\mathrm{w}}} f^{\prime}(\omega_0 / k_{\mathrm{w}})~,
\end{equation}
implying that the equilibrium is unstable for positive $f^{\prime}(\omega_0 / k_{\mathrm{w}})$ 
whereas perturbations are damped for negative $f^{\prime}(\omega_0 / k_{\mathrm{w}})$. 

For equally spaced beams, $v_s - v_{s'} = (s-s') \Delta p$, 
analytical estimates \citep{Dawson1960,EscZekEls1996,EEbook2003} 
show that a pair of eigenvalues $\omega_r \pm \rmi \gamma_r$ 
is located near each beam (with $\vert v_s - \omega_r / k_{\mathrm{w}} \vert < \Delta p / 2$),
with imaginary part 
\begin{equation}
  \gamma_r 
  \simeq 
  - \frac 1 {\tau_{\mathrm{rec}} }
     \ln \Bigl| \frac {\delta \sqrt{1 + \theta_r^2}} {4 \pi} \Bigr| 
  + \ldots 
  \label{ch3gammAppr}
\end{equation}
where $\delta := f'(v_s) \Delta p / f(v_s)$ and $\theta_r = {\mathrm{O}}(1)$. 
Their real part $\omega_r$ is near $( v_s + (\Delta p/4) \, {\mathrm{sign}}\,\delta ) k_{\mathrm{w}}$. 
The recurrence time
\begin{equation}
  \tau_{\mathrm{rec}} 
  = 
  \frac {2 \pi} {k_j \Delta p}
\label{SZ117a}
\end{equation}
characterizes the time scale on which the approximation of 
the many beams by a smooth distribution function loses its validity.

In the continuum limit, the imaginary parts $\pm \gamma_r$ of these eigenfrequencies 
as well as their spacing go to zero, 
so they approach a continuum spectrum of real eigenfrequencies 
which corresponds to the analogue of van Kampen modes found in the vlasovian approach. 

When the distribution function has a negative slope, there are $2b$ such van Kampen-like eigenvalues,
and just one real eigenvalue beyond the fastest beam (see figure \ref{fig:dispersion-relation}). 
In contrast, for a positive $f'(\omega_0 / k_{\mathrm{w}})$, 
besides these solutions condensing to the real axis in the limit,
equation \eqref{eq:dispersion-relation} is also satisfied by a particular pair of eigenvalues 
$\omega_{\mathrm{L}} \pm \rmi \gamma_{\mathrm{L}}$
whose imaginary part is given (in the limit $\Delta p \to 0$) by \eqref{eq:Landau-growth-rate}.

A first benefit of the discretized model with monokinetic beams is 
its avoiding singular integrals and prescriptions of proper integration contour inherent to the kinetic approach.
It also stresses a deep difference between the damping and growth cases :
despite their being encompassed by a single formula \eqref{eq:Landau-growth-rate},
they must involve different physics, because the instability is associated with a single eigenvalue solving \eqref{eq:dispersion-relation}, 
while in the damping case equation \eqref{eq:dispersion-relation} admits \emph{no solution} with negative imaginary part near $\gamma_{\mathrm{L}}$. 

This paradox reflects the reversibility of hamiltonian dynamics~: 
if damping resulted from a genuine eigenmode of the system, 
the dispersion relation would have a conjugate eigenvalue with positive real part, leading to an instability. 
This paradox of the Vlasov equation was solved by \citet{vanKampen1955, vanKampen1957} and \citet{Case1959}.

A second indication that Landau damping does not result from the ``dominant eigenmode'' in kinetic theory 
is that some initial data lead to a damping with smaller decay rate than Landau's~:
when present, they dominate over the Landau behaviour for long times \citep{Belmont2011}.

Actually, we shall see that 
both Landau damping and growth are related to a phase mixing effect among the van Kampen-like modes 
when the general solution of the linearized system is written in terms of its eigenfunctions.


\subsection{Normal modes expansion}
\label{sec:normal modes}

Denote by $\cC_r$ and $\sigma_r$ the $r$-th eigenvector of $\cM$ and its eigenvalue. 
Then, by the linearity of equation \eqref{eq:linear-system}, the general solution $\cG(t) = [C_1(t) \dots C_b(t),A_1(t) \cdots, A_b(t),Z(t)]^{\top}$ is a 
superposition of the eigensolutions $\cC_r \, \rme^{-\rmi \sigma_r t}$, i.e.\ 
\begin{equation}
	\label{eq:mode-expansion}
	\cG(t) = \summ_{r = 1}^{2b + 1} \xi_r \, \cC_r \, \rme^{-\rmi \sigma_r t}~,
\end{equation}
\noindent with components 
\begin{eqnarray}
	\label{eq:mode-expansion-comp_Cs}	
	C_s(t) &=& \frac{\varepsilon}{2} \summ_{r = 1}^{2b + 1} \frac{\xi_r}{(\sigma_r - k_{\mathrm{w}} v_s)^2} \, \rme^{-\rmi \sigma_r t}~,\\
	\label{eq:mode-expansion-comp_As} 
	A_s(t) &=& \frac{\varepsilon}{2} \summ_{r = 1}^{2b + 1} \frac{\xi_r}{\sigma_r - k_{\mathrm{w}} v_s} \, \rme^{-\rmi \sigma_r t}~,\\
	\label{eq:mode-expansion-comp_Z} Z(t) &=& \summ_{r = 1}^{2b + 1} \xi_r \ \rme^{-\rmi \sigma_r t}~.
\end{eqnarray}
The coefficients $\xi_r$ of the linear combination are obtained from the initial condition and the left eigenvectors of matrix $\cM$, $\sigma_r \cC^{\prime}_r = \cM^{\top} \cdot \cC^{\prime}_r$, 
through the inner product
\begin{equation}
	\label{eq:xis}
	\xi_r = \cC^{\prime \top}_r \cdot \cG(0) = z^{\prime}_r  \left[ Z(0) + \varepsilon \summ_{s = 1}^{b} N_s \left(\frac{C_s(0)}{\sigma_r - k_{\mathrm{w}} v_s} -\frac{A_s(0)}{(\sigma_r - k_{\mathrm{w}} v_s)^2} \right)\right]~,
\end{equation}
\noindent where the scaling factor $z^{\prime}_r$, given by
\begin{equation}
	\label{eq:scaling-factor}	
	z^{\prime}_r = \left[1 + \varepsilon^2 \summ_{s = 1}^{b} \frac{N_s}{(\sigma_r - k_{\mathrm{w}} v_s)^3}\right]^{-1}~,
\end{equation}
\noindent is chosen so that the normalization condition $\cC^{\prime\top}_{r'} \cdot \cC_r = \delta_{r',r}$ is satisfied.

In the linear regime, the system of beams interacting with a single Langmuir wave 
is therefore analytically solvable in terms of a normal modes expansion of the linearized equations of motion. 
Within the present approach, 
specifically through equations \eqref{eq:dx-Fourier2}-\eqref{eq:A-ball} and \eqref{eq:mode-expansion}-\eqref{eq:scaling-factor}, 
note that once the full spectrum of eigenfrequencies $\{\sigma_r\}$ and the initial condition of the dynamics 
(the coefficients $\xi_r$'s) are known,   
the evolution of the wave amplitude and the perturbations on the particles orbits are determined.

In particular, the scaling factor $z'_r$ coincides with the weight of eigenvector $\cC_r$ 
in the decomposition of a ``quiet start'' initial condition 
($\delta x_{ns}(0) = 0, \delta p_{ns}(0) = 0$, viz.\ $C_s(0) = 0, A_s(0) = 0$). 


\section{Numerical results and discussion}
\label{sec:Results}

In this section, we discuss the spectrum of the van Kampen-like eigenfrequencies obtained by solving numerically equation \eqref{eq:dispersion-relation} for many-beam systems and stress some microscopic aspect of the dynamics. 
As we shall see, \emph{both} wave damping and growth arise as consequence of an interference 
among the $2b + 1$ eigenmodes. 
However, in respect to this interference (or phase mixing) effect, 
these two cases exhibit a remarkable distinctive behaviour as the number of beams gets large.

The monokinetic beams approach presented in section \ref{sec:Formalism} breaks down 
after times of the order of bouncing time $\omega^{-1}_{\mathrm{b}} = (\varepsilon |Z| k_{\mathrm{w}}) ^{-1/2}$, 
where the particles orbit may no longer be considered approximately ballistic. 
To ensure that trapping effects play a negligible role and the structure of beams is preserved, 
all the following results were obtained in the weak amplitude regime 
(easily implemented in taking $\varepsilon \ll 1$, $N \sim \varepsilon^{-2} \to \infty$). 
To simplify our numerical calculations, we take $\omega_0 = 0.0$ which, from the physical viewpoint, 
amounts to performing a Galileo transformation to a reference frame moving with the wave (nominal) phase velocity. 
We set the equilibrium state with particle velocities equally spaced ($v_s = v_1 + (s - 1) \Delta p$)
inside an interval of length $I_v$ centered at the origin and the distribution function $f(v_s) = f(v_1) + (s - 1) f^{\prime} \Delta p$, 
where $f^{\prime}$ is constant and $\Delta p = I_v / b$ is the velocity interval between two adjacent beams. 
The data $f(v_1) = 1/I_v - (b - 1) \Delta p f^{\prime} / 2$ ensures that 
the distribution of velocities is normalized to unity. 
Thus, the number of particles \cite{note2} 
in beam $s$ is given by $N_s = N \Delta p f (v_s)$.

\begin{figure}
	\centering
		\includegraphics[width = 0.85\linewidth]{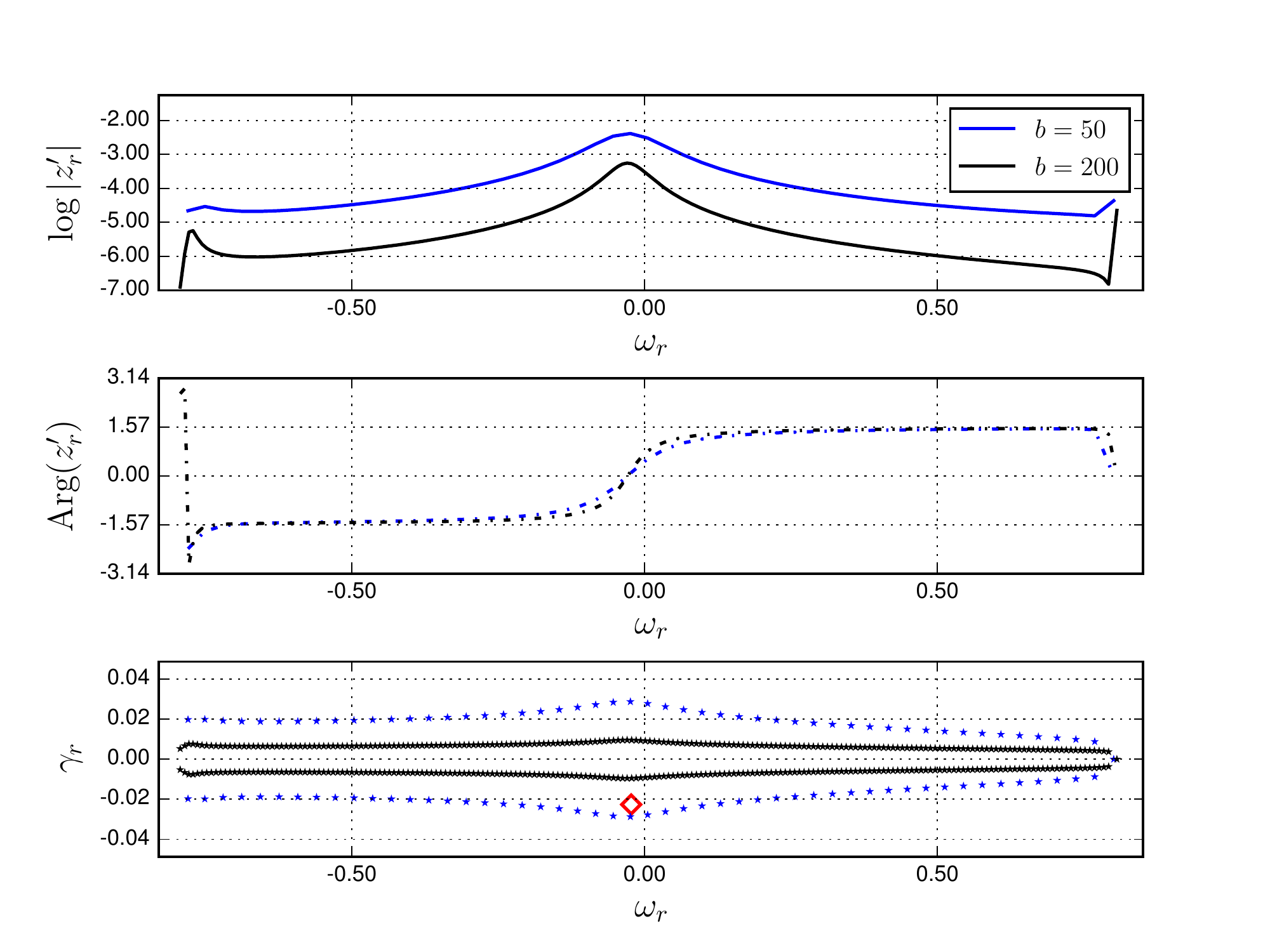}\\
		\caption{Damping case : discrete analogue of van Kampen eigenfrequencies in the complex $\sigma$ - plane (bottom), the phase (center) and the modulus (top) of the scaling factor. 
		The curves in the two upper figures are guides to the eye.  
		The red losangle locates the Landau value $\sigma_{\mathrm{L}}$ (which corresponds to no eigenmode in the finite-$N$ model). 
		The results were obtained for systems composed of $50$ and $200$ beams with parameters set to $I_v = 1.6$, $\omega_0 = 0.0$, $k_{\mathrm{w}} = 1.0$, $\varepsilon N f^{\prime} = -4.8$ and $\gamma_{\mathrm{L}} = -2.26 \times 10^{-2}$.}\label{fig:spectrum_ST}
\end{figure}

In the weak-coupling regime and for many-beam systems, the spectrum of eigenfrequencies is very close to the real axis, 
and therefore the numerical solutions of equation \eqref{eq:dispersion-relation} must be computed with high accuracy. 
To obtain the spectrum of eigenfrequencies for such systems 
with an accuracy down to $10^{-7}$ ($ \vert \sigma^{\mathrm{approx}}_r - \sigma_r \vert \leq 10^{-7}$), 
we developed a root-finding scheme based on Cauchy's residue theorem 
with triangular contours that enabled us to implement a bisection method in the complex $\sigma$-plane. 
In this paper, we shall not dwell on the technical details of the method~;
its key ideas are discussed in appendix \ref{sec:AppA}.


\subsection{Spectra and scaling factors}
\label{sec:spectra}

\begin{figure}	
	\centering
		\includegraphics[width = 0.85\linewidth]{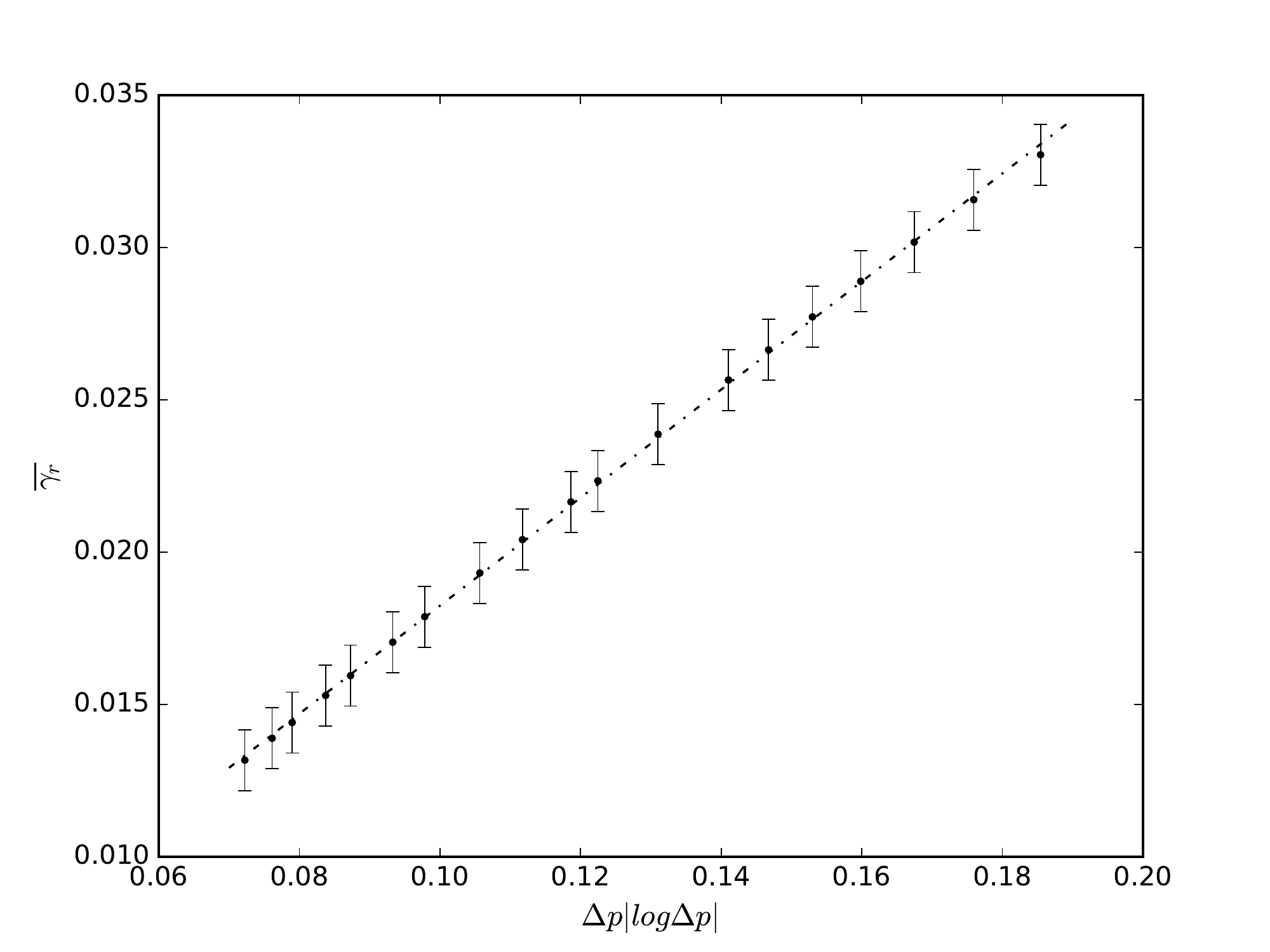}
		\caption{Damping case : behaviour of the average distance to real axis 
		$\bar{\gamma_r} \equiv b^{-1} \sum_{r, \gamma_r > 0} \gamma_r$ as a function of the beam spacing. 
		The linear regression $y = (0.177 \pm 0.001)x \pm 0.001$ shows that $\gamma_r$ goes to zero 
		like $x \equiv \Delta p \ \vert \log \Delta p \, \vert$ as computed by \citet{Dawson1960}. 
		The error bars represent the tolerance $10^{-3}$ used in the root-finding routine.}
\label{fig:asymp}
\end{figure}

In the lowest graph in figure \ref{fig:spectrum_ST}, we plot the spectrum of van Kampen-like eigenfrequencies for the damping case ($f^{\prime} < 0$) for systems composed of $50$ and $200$ beams. 
The upper two plots display the logarithm of the modulus of the scaling factor $z^{\prime}_r$
and its phase ${\mathrm{Arg}} \, z^{\prime}_r$, obtained from the spectra and \eqref{eq:scaling-factor}. 
These curves correspond only to the eigenmodes with $\gamma_r \geq 0$~: 
thanks to time-reversibility of hamiltonian dynamics implying real coefficients in the algebraic equation \eqref{eq:scaling-factor}, 
the stable eigenmodes differ only by the opposite sign of their phase.
Regardless of the number of beams, 
the phase velocities $\omega_r/k_{\mathrm{w}}$ of the damped and growing eigenmodes 
are confined to the interval $[-0.8,0.8]$ that contains the resonant particles. 
The more beams we consider, denser the spectrum becomes and it approaches as a whole the real axis. 
The dependence of the mean value of the imaginary part $\bar{\gamma_r} = b^{-1} \sum_{r, \gamma_r > 0} \gamma_r$ 
(which amounts to the $\ell^1$ norm of the sequence of these imaginary parts) on the beam spacing 
is illustrated in Figure \ref{fig:asymp}, indicating how, 
in the continuous limit ($\Delta p \rightarrow 0$), these eigenfrequencies accumulate toward the real axis 
forming what corresponds to the continuous spectrum of van Kampen frequencies derived from a vlasovian description. 
The spacing between the real parts of successive eigenfrequencies also goes to 0 like $\Delta p$ 
because there is always a real $\omega_r/k_{\mathrm{w}}$ between two successive beam velocities,
so that the condensation of eigenvalues onto the real axis tends to a cut along the interval $[\inf(v_s), \sup(v_s)]$~:
this cut is the locus of van Kampen's singular spectrum. 
This behaviour $\gamma_r \sim \Delta p \, \vert \ln \Delta p \, \vert $, \eqref{ch3gammAppr} 
was computed by \citet{Dawson1960}.

Landau's value $\sigma_{\mathrm{L}} = \omega_{\mathrm{L}} + \rmi \gamma_{\mathrm{L}}$, 
represented by the red diamond point in figure \ref{fig:spectrum_ST}, 
is obtained after taking the continuous limit of equation \eqref{eq:dispersion-relation}. 
The interaction between the wave and resonant particles is responsible 
for the small deviation $\vert \omega_{\mathrm{L}} - \omega_0 \vert = 2.30 \times 10^{-2}$ 
from the free wave frequency, which can also be estimated analytically \citep{EEbook2003}.

We also observe, in the vicinity of $\omega_{\mathrm{L}}$, 
a striking behaviour for the modulus and for the phase of the scaling factor. 
The top plot evidences a peak slightly shifted from the origin, 
and the center plot shows a change in the phase's sign. 
These two behaviours together indicate 
that the modes with greater contribution to the initial data $(\delta x_{ns}(0), \delta p_{ns}(0),Z(0))$ 
are those with a frequency close to the Landau value,
in agreement with the estimate \citep{EEbook2003} 
\begin{equation}
  z'_r
  \simeq
  -  \frac{1}{2 \pi} \frac {k_{\mathrm{w}} \Delta p} {\gamma_{\mathrm{L}} \pm \rmi (\omega_r - \omega_{\mathrm{L}})}
  \label{ch3scalzvK2}
\end{equation}
from \eqref{eq:scaling-factor}. The plus-or-minus sign in this equation take care of the two $r$-eigenmodes with same phase velocity.

\begin{figure}
	\centering
	\begin{minipage}[t]{.85\linewidth}
		\centering
		\includegraphics[width = \linewidth]{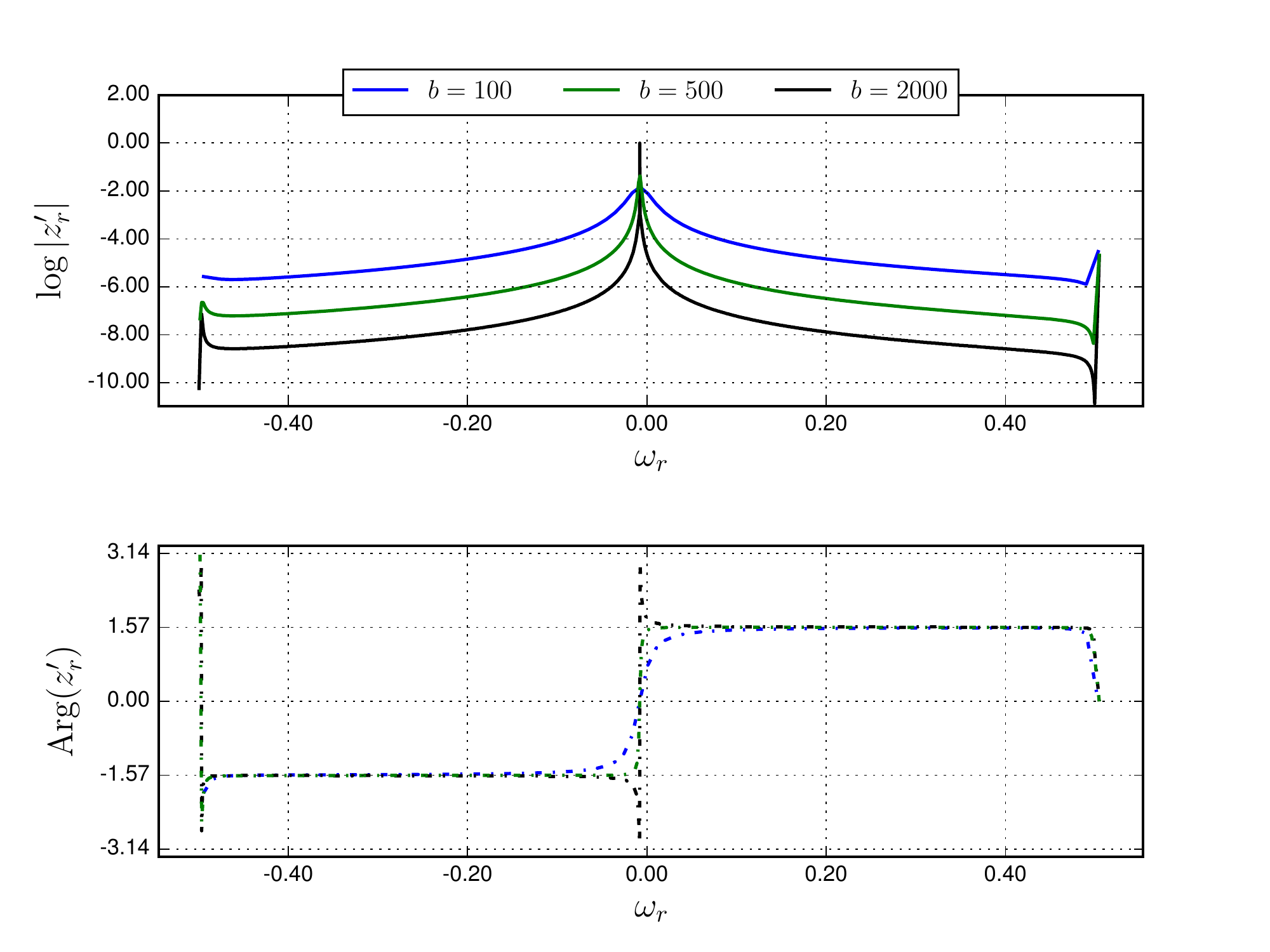}
		\label{fig:scaling-factor-UNST-a}
	\end{minipage}
	\begin{minipage}[t]{.85\linewidth}
		\centering
		\includegraphics[width = \linewidth]{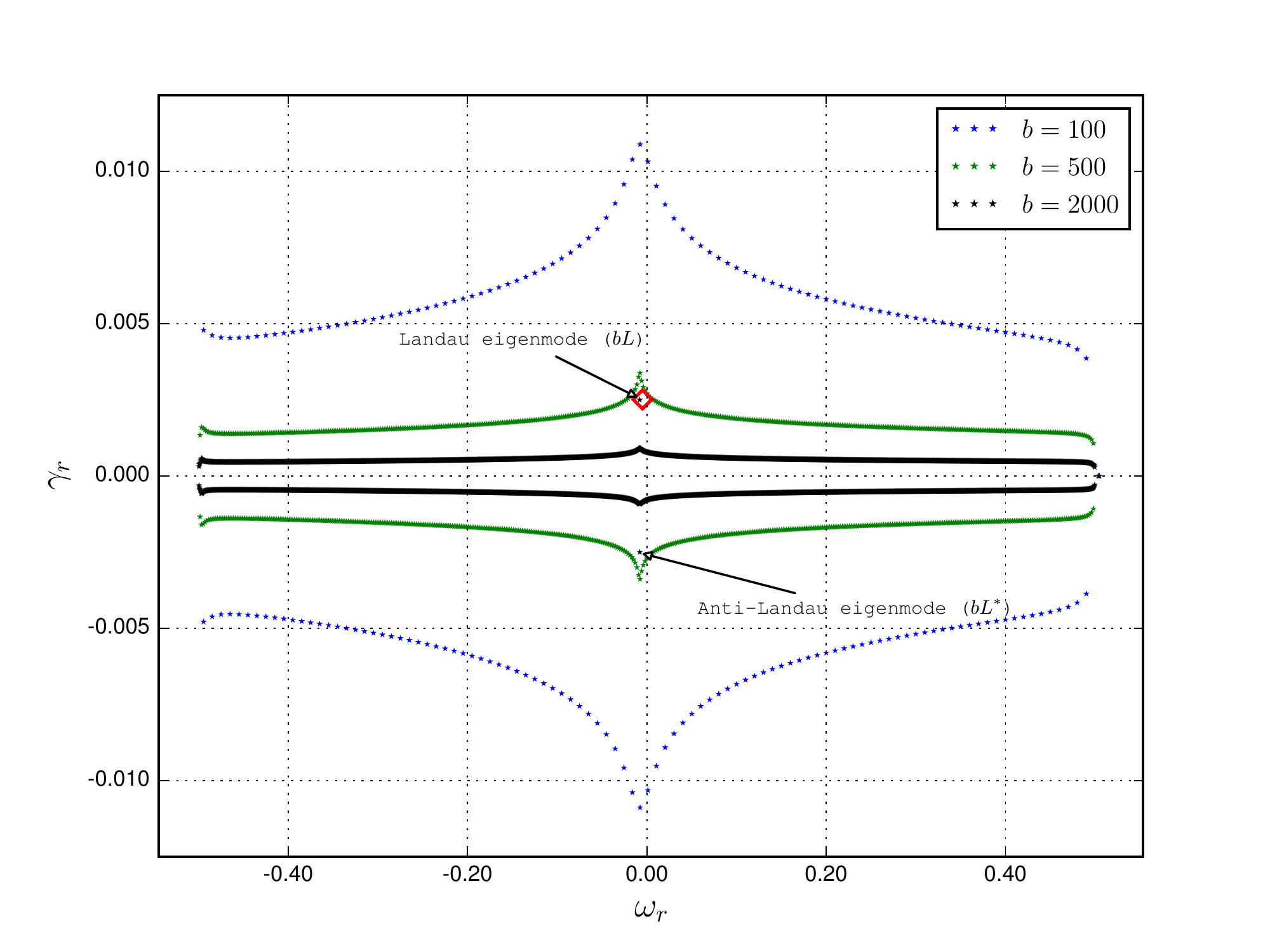}\\ 
		\label{fig:spectrum-UNST-b}	
	\end{minipage}
		\caption{Growth case : (top and middle) modulus and phase of the scaling factor (the lines are guides to the eye), and
		(bottom) spectrum of the van Kampen-like eigenfrequencies in the complex $\sigma$ plane for systems composed of $100$, $500$ and $2000$ beams. 
		The red losangle locates the Landau mode, which is slightly shifted from the origin by $\omega_{\mathrm{L}} = -4.87 \times 10^{-3}$.
		The parameters were set to $I_v = 1.0$, $\omega_0 = 0.0$, $k_{\mathrm{w}} = 1.0$, $\varepsilon N f^{\prime} = 16.0$ and $\gamma_{\mathrm{L}} = 2.51 \times 10^{-3}$.}
 \label{fig:spectrum_UNST}
\end{figure}

For the growth case ($f^{\prime} > 0$), the spectrum and the scaling factor shown in figure \ref{fig:spectrum_UNST} 
exhibit two distinctive features (compared to the damping regime) for systems with many beams. 
The first one is the presence of two specific eigenfrequencies, complex conjugate to each other, 
that do not approach the real axis as the number of beams increase. 
We denote the eigenfrequency close to the Landau frequency by $\sigma_{b{\mathrm{L}}}$ (Landau-like eigenmode, with subscript $b$ recalling the finite number of beams) 
and its complex conjugate by $\sigma_{b{\mathrm{L}}^{\ast}}$ (anti-Landau eigenmode), 
both highlighted in the figure. 
The second striking difference is the spike (mind the log scale) in the curve for the modulus of the scaling factor. 
It shows, along with ${\mathrm{Arg}}(z'_{b{\mathrm{L}}}) \approx 0$, that the Landau and the anti-Landau eigenmodes have a dominant contribution to the initial data.


\subsection{Quiet start evolution}
\label{sec:quiet-start}

In order to monitor the contributions of the $b{\mathrm{L}}$ and $b{\mathrm{L}}^{\ast}$ eigenmodes to the wave amplitude, 
we consider a single realization of the system where the wave is launched with initial amplitude $Z(0) = 1$ 
and particles start from the equilibrium configuration of monokinetic arrays. 
For this specific quiet start realization, the evolution of the wave amplitude is given according to equation \eqref{eq:mode-expansion-comp_Z} by
\begin{equation} \label{eq:Z_decomposed}
  Z(t) 
  = 
  z^{\prime}_{b{\mathrm{L}}} \ \rme^{-\rmi \sigma_{b{\mathrm{L}}} t} 
  + z^{\prime}_{b{\mathrm{L}}^{\ast}} \ \rme^{-\rmi \sigma_{b{\mathrm{L}}^\ast} t} 
  + \summ_{\substack{r = 1  \\   r \neq b{\mathrm{L}},b{\mathrm{L}}^{\ast}}}^{2b + 1} z^{\prime}_r \ \rme^{-\rmi \sigma_r t}~.
\end{equation}
recalling that the scaling factor corresponds to a weight factor indicating how much each eigenmode contributes to construct the initial amplitude.
\begin{figure}
	\centering
		\includegraphics[width = 0.85\linewidth]{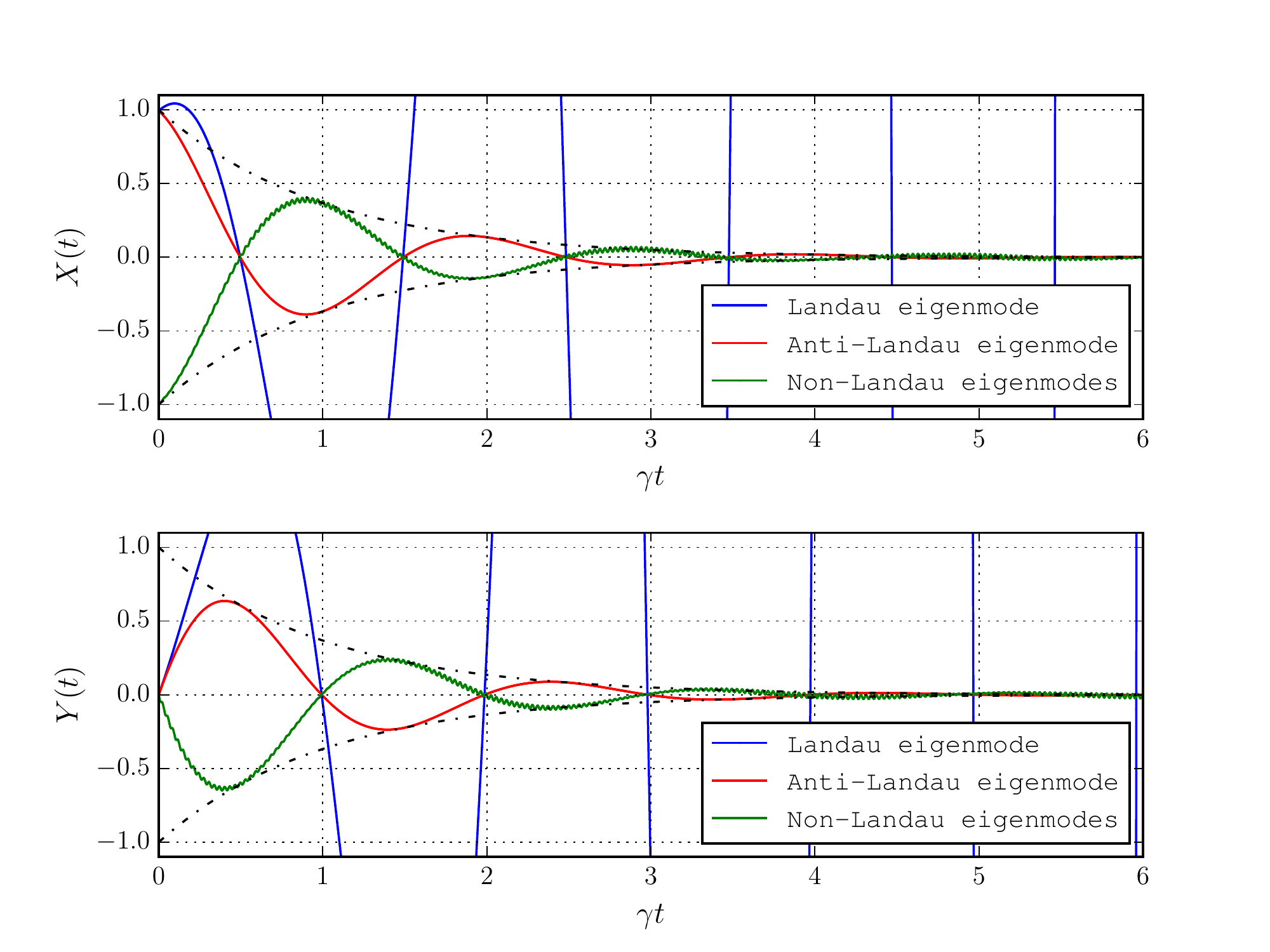}
		\caption{Growth case : evolution of real (top) and imaginary (bottom) parts of the Landau ($b{\mathrm{L}}$), anti-Landau ($b{\mathrm{L}}^{\ast}$) and van Kampen-like components in the wave complex amplitude $Z = X + \rmi Y$ for a system of 2000 beams in the unstable case. 
		The symmetry between the red and green curves shows the compensation between the anti-Landau eigenmode and the van Kampen spectrum.
		The Landau (blue) line is growing exponentially, exceeding largely the range of our ordinate axis.} \label{fig:cancellation}
\end{figure}

Due to the fact that for many-beam systems $z^{\prime}_{b{\mathrm{L}}} \approx 1$ 
and $\sigma_{b{\mathrm{L}}} \approx \sigma_{\mathrm{L}}$, 
as seen from the black graphs in figure \ref{fig:spectrum_UNST}, 
one would expect the Landau-like ($b{\mathrm{L}}$) eigenmode to provide, by itself, the correct growth of the wave. 
However, the contribution of the other $2b$ eigenmodes still have to be taken into account. 
Thanks to a destructive interference, these $2b$ eigenmodes superposed contribute only with a small oscillatory part that does not compromise the exponential growth $\sim \rme^{\gamma_{\mathrm{L}} t}$ of the wave. 
Figures \ref{fig:cancellation} illustrate this destructive interference 
by monitoring the evolution of the cartesian components of each term of equation \eqref{eq:Z_decomposed} 
for a system of $2000$ beams. 
This graph exhibits a clear symmetry between the red and green curves, 
showing that the O(1) contribution of the anti-Landau eigenmode is cancelled out by the superposition of the (dense, individually small) van Kampen-like eigenmodes. 
The dashed curves that appear as an envelope in Fig. \ref{fig:cancellation} 
are given by $\pm \rme^{-\gamma_{\mathrm{L}} t}$ and indicate that, 
in the same way as for the anti-Landau eigenmode, 
the superposition of the van Kampen-like eigenmodes also damps with the anti-Landau rate.

Analytically, indeed, the asymptotic form \eqref{ch3scalzvK2} for the scaling factor implies 
that the third term in \eqref{eq:Z_decomposed} behaves like the Fourier transform of a Lorentz function, 
namely as $ - Z(0) \, \rme^{- \gamma_{\mathrm{L}} \, \vert t \vert}$. 
Therefore, \eqref{eq:Z_decomposed} reduces to 
\begin{equation}
  Z(t) 
  \simeq 
  Z(0) \, \rme^{\gamma_{\mathrm{L}}  t } 
  + Z(0) \, \rme^{- \gamma_{\mathrm{L}}  t }
  - Z(0) \, \rme^{- \gamma_{\mathrm{L}} \, \vert t \vert}
  \label{eq:Z-1}
\end{equation}
which reduces to $Z(0) \, \rme^{\gamma_{\mathrm{L}}  \, \vert t \vert}$ for both positive and negative $t$. 

This form rescues the time-reversibility in the solution to the initial-value problem. 
If the response to the initial perturbation were described with only the Landau-like eigenmode, 
it would lead to a decreasing $\vert Z(t) \vert$ for $t \to - \infty$. 
Yet time-reversal symmetry (unbroken by the quiet start initial condition) requires 
the wave to increase as $t \to - \infty$ just as it does for $t \to \infty$. 
Therefore, the system evolution for all times must involve both 
the Landau-like and anti-Landau modes, the sum of which reads $2 Z(0) \cosh \gamma_{\mathrm{L}} t$. 
But this sum does not start exactly exponentially at $t = 0$, 
so that the van Kampen-like eigenmodes are needed to cancel the anti-Landau mode for $t>0$,
and to cancel the Landau-like mode for $t<0$. 

This peculiarity of the growth regime can be seen numerically only when considering a larger number of beams. 
The necessity of working with such many-beam systems motivated us to developed the Cauchy root-finding scheme once that traditional computer algebra systems failed in providing the full spectrum. 

Note that the damping case is analytically simpler~:
it involves only van Kampen-like eigenmodes, for which the scaling factors given by \eqref{ch3scalzvK2} 
again yield the Fourier transform of a Lorentz function. 
However, $\gamma_{\mathrm{L}}$ is now negative, so that $z'_r > 0$ for $\omega_r \approx \omega_L$
and \eqref{eq:Z_decomposed} reduces to
the well known 
\begin{equation}
  Z(t) \simeq Z(0) \, \rme^{\gamma_{\mathrm{L}}  \, \vert t \vert} \, .
  \label{eq:Zdecay}
\end{equation}

\begin{figure}	
	\centering
		\includegraphics[width = 0.85\linewidth]{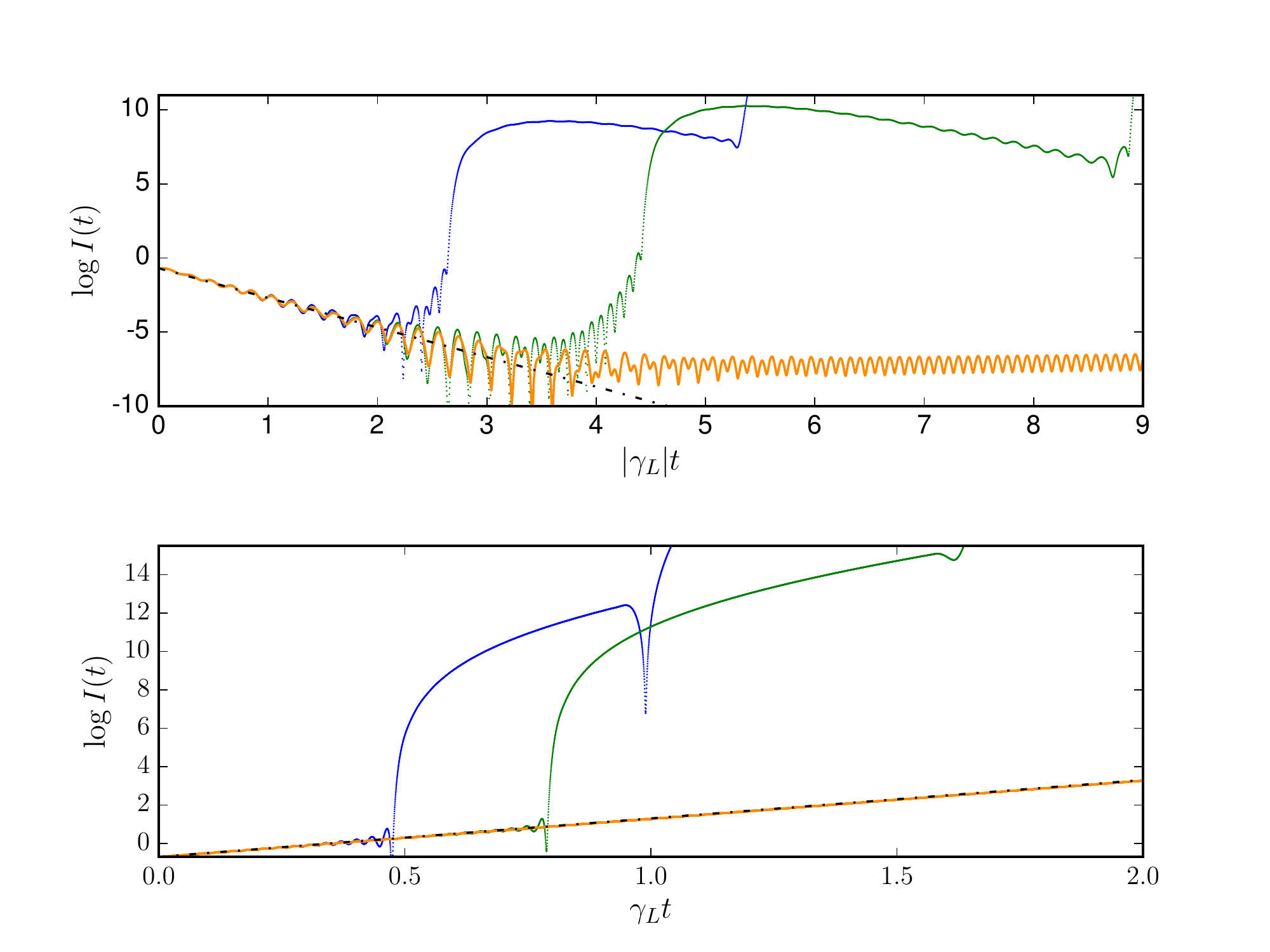}
	\caption{Evolution of the wave intensity for systems composed of $30$ (red), $50$ (blue) and $200$ (orange) beams for the stable (top) and the unstable (bottom) cases with the dashed lines corresponding to the Landau prediction.
	The same parameters that generated the spectra in figures \ref{fig:spectrum_ST} and \ref{fig:spectrum_UNST} were used here. 
	As initial condition we consider $Z(0) = 1.0$ and unperturbed trajectories $\delta x_{ns}(0) = \delta p_{ns}(0) = 0$ with particles starting in the configuration of equally spaced monokinetic beams.}\label{fig:evolution}
\end{figure}

Figures \ref{fig:evolution} show the evolution of the wave intensity $I = Z^{\ast} Z/2$, obtained through the superposition of the van Kampen-like eigenmodes for systems composed of $30$, $50$ and $200$ beams. 
We observe that the Langmuir wave damps (or grows) initially with the expected Landau prediction $\log I_{\mathrm{L}}(t) = \log I(0) \pm 2 \vert \gamma_{\mathrm{L}} \vert t$, represented by the dashed black lines. 
For the unstable case, even for a small number of beams, 
where $\max\{\gamma_r\} > \gamma_{\mathrm{L}}$ and the $b{\mathrm{L}}$ and $b{\mathrm{L}}^{\ast}$ modes are not prominent, 
the discrete and continuous systems agree.

The divergences observed in the graphs occur for times close to the recurrence time 
$\tau_{\mathrm{rec}} \sim 2 \pi / \Delta \omega_r \sim 2 \pi / (k_{\mathrm{w}} \Delta p)$ 
where the lack of an effective phase mixing of the van Kampen-like eigenmodes 
along with the large values of the modulus of the complex amplitude $z^{\prime}_r \ \rme^{\gamma_r t}$ of the unstable eigenmodes 
make the wave intensity depart from the Landau line. 
After this characteristic time, the approximation of  a discrete system composed of monokinetic beams 
by a continuous one is no longer valid \citep{Firpo1998}. 
It is worth noting that the departure of the orange curve from the Landau line in the top part of Figure~\ref{fig:evolution} 
is not related to the breakdown of the phase mixing but to the small values of $I_{\mathrm{L}}(t)$ compared to the amplitude of the oscillations.

\section{Summary and perspectives}
\label{sec:Conclusion}

Our results in this paper provide an accurate numerical support for studying the phase mixing mechanism in hamiltonian models. 
Success in investigating this process for many-beams systems rests on the development of a new root finding scheme based on the Cauchy's integral theorem enabling one to compute all the roots of the dispersion relation \eqref{eq:dispersion-relation}.

Analysing the spectrum of eigenfrequencies of the van Kampen-like modes and the evolution of the cartesian components of the wave amplitude
highlights a remarkable aspect in which the stable and unstable regimes differ. 
In the stable case, the normal modes simply contribute collectively in providing the Landau damping~;
however, in the unstable case, the \emph{pure} exponential growth of the wave amplitude 
is the result of a destructive interference effect 
(between its dual exponentially decaying mode, with same initial amplitude, 
and a flurry of van Kampen modes with small individual amplitudes) 
that enables a single eigenmode to provide a dominant contribution on the dynamics from the very beginning. 
We conclude the paper pointing out the consistence between the discrete and the continuous systems \citep{Firpo1998}, 
and verifying that the superposition of the van Kampen-like eigenmodes indeed provides the expected exponential Landau damping/growth rate for the wave intensity.

The quiet start as well as the equally spaced beams setting were chosen in order to simplify the calculations.
However, one expects similar results for other discretizations and initial conditions of the system in the many beams regime. 

The first author thanks Coordena\c{c}\~ao de Aperfei\c{c}oamento de Pessoal de N\'{i}vel Superior (CAPES) 
for financing his stay at Aix-Marseille Universit\'{e} under the Programa de Doutorado Sandu\'{\i}che no Exterior (PDSE), 
process No. $99999.004512/2014-06$.
Valuable discussions with BV Ribeiro and comments from DF Escande, MA Amato, F Doveil and DFG Minenna
are gratefully acknowledged.
The authors also aknowledge the reviewers for their useful comments.

\appendix
\section{Bisection-like root finding method in the complex plane}
\label{sec:AppA}

\begin{figure}
	\centering
	\begin{minipage}[t]{.49\linewidth}
		\centering
		\includegraphics[width = \linewidth]{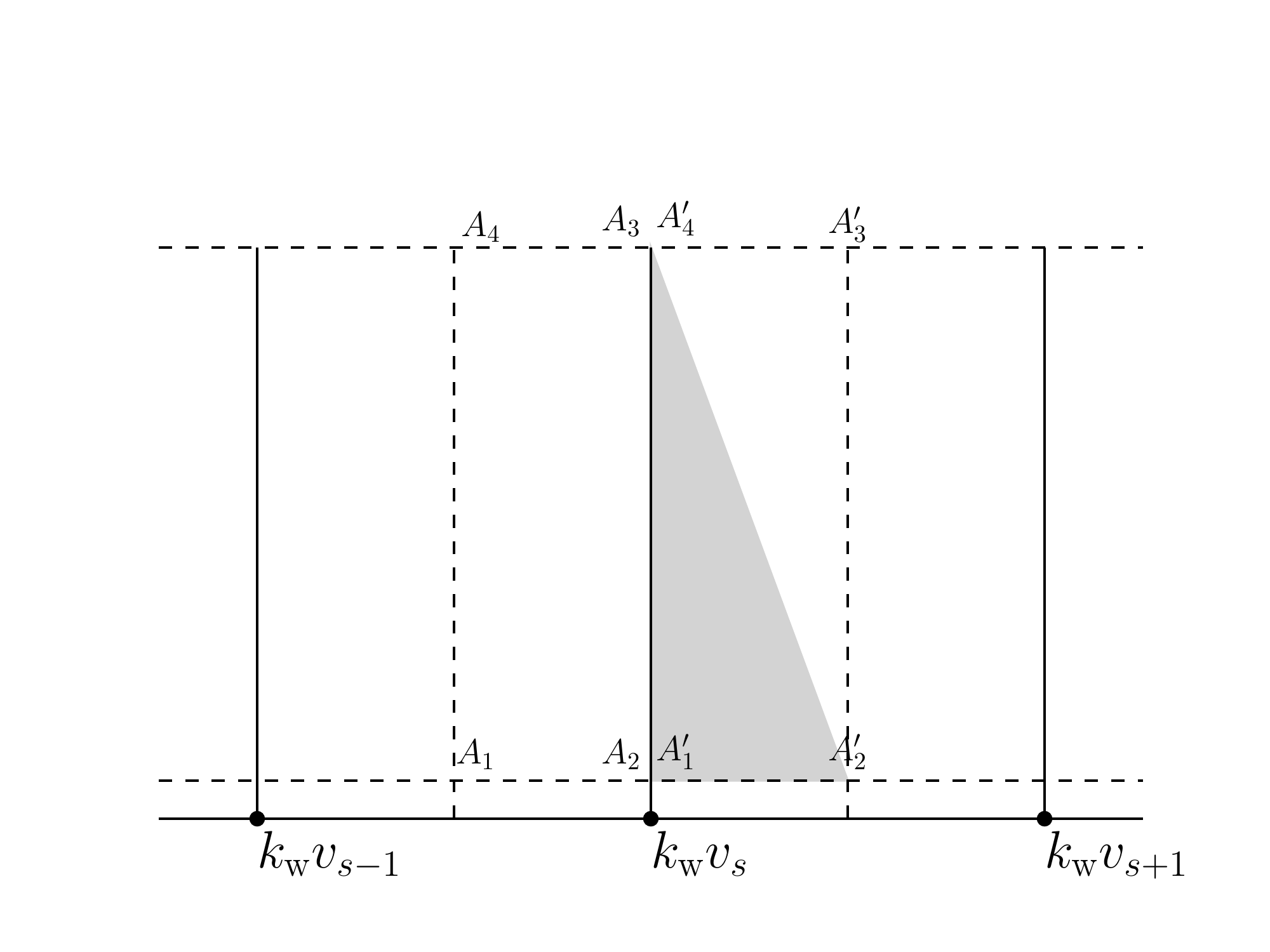}
	\end{minipage}%
	\begin{minipage}[t]{.49\linewidth}
		\centering
		\includegraphics[width = \linewidth]{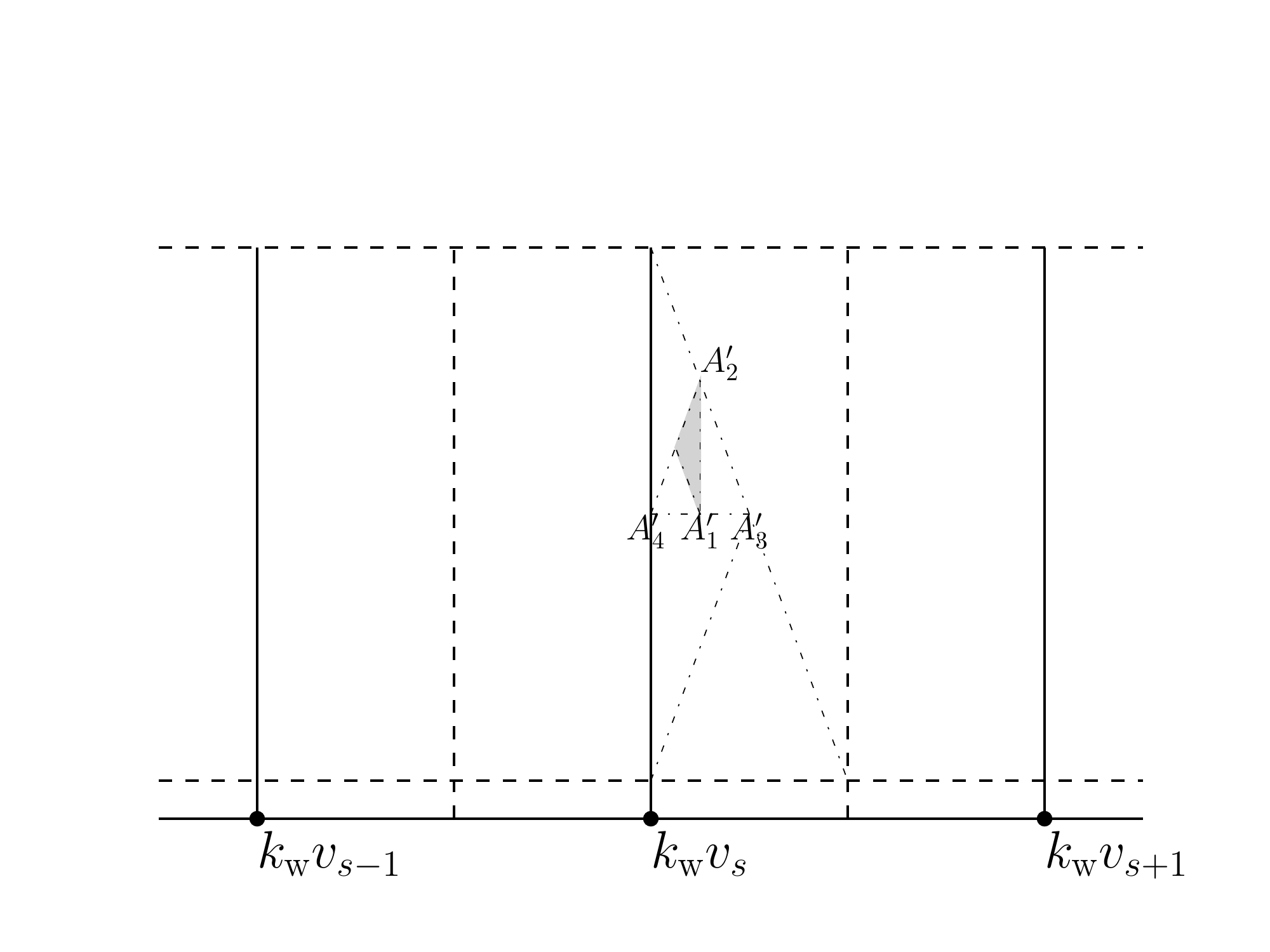}\\ 
	\end{minipage}
	\caption{Illustration of the bisection-like method in the complex $\sigma$-plane by supposing that the root is located to the right side of the line corresponding to the $s$-th beam. 
The shaded triangles highlight the regions for which the contour integral is nonzero and that, consequently, contain a root inside.}
\label{fig:root-finding}	
\end{figure}

\begin{figure*}
		\includegraphics[width = 0.5\linewidth]{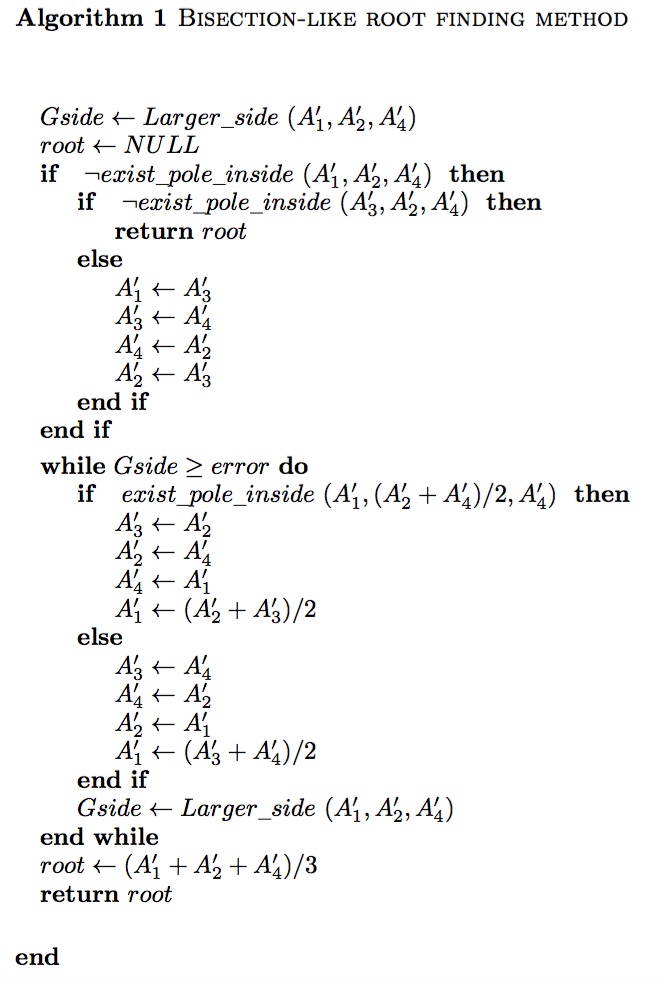}
\end{figure*}

The Cauchy integral theorem states that if $F$ is an analytic function on a simply connected domain $\cD$ 
bounded by a simple closed contour $\partial \cD$, then $\oint_{\partial D} F(z)  \rmd z = 0$, 
and if $F$ has a finite number of simple poles $z_j$ in $\cD$ with residues $R_j$, 
then $\oint_{\partial D} F(z)  \rmd z = 2 \pi \rmi \sum_j R_j$.
In this appendix, we expose a direct scheme based on this elementary tool of complex analysis 
with the aim of computing numerically the full set of van Kampen-like eigenfrequencies.

With $\chi(\sigma)$ being the right rand side of equation \eqref{eq:dispersion-relation}, 
the method consists in obtaining the root of $\sigma -\chi(\sigma)$, inside a given region, 
by searching for the pole of $(\sigma -\chi(\sigma))^{-1}$. 
To find a root in the vicinity of the velocity of the $s$-th beam, we define two rectangles~: 
the left one with vertices $A_1$ to $A_4$ and the right one with vertices $A^{\prime}_1$ to $A^{\prime}_4$, 
both illustrated in Fig.~\ref{fig:root-finding}(left). 
Firstly, we estimate the Cauchy integral along these rectangular contours in order to identify 
to which side from the $s$-th beam the root is located. If one of these rectangles yields
\begin{equation}
\left\vert \oint_{\partial \cD} (\sigma - \chi(\sigma))^{-1} d\sigma \right\vert \leq C_{\mathrm{crit}},
\label{eq:Cauchy-criterion}
\end{equation}
\noindent with $C_{\mathrm{crit}}$ representing the Cauchy criterion to assume the integral numerically zero, it is discarded as a possible region containing the root.

After identifying the presence of a root inside a rectangle, the next step consists in implementing the bisection-like method 
which is based on successive divisions of the region of search. 
In figure \ref{fig:root-finding}, we illustrate this procedure with the shaded triangles in figures \ref{fig:root-finding}(left) and \ref{fig:root-finding}(right) 
showing, respectively, the configuration at the beginning and after $6$ iterations of the method. 
Iterations are stopped when the largest side of the shaded triangle is smaller than the tolerance of the method (which is an input data). 
Following so, the approximate root of $\sigma - \chi(\sigma)$ is given by the geometrical center of the triangle.

The bisection-like method is displayed through the pseudo-code in Algorithm 1, 
where we called the function ``Larger\underline{ }side (A,B,C)", 
that yields the length of the largest side of triangle ABC, 
and the function ``exist\underline{ }pole\underline{ }inside (A,B,C)" whose boolean output informs, 
by means of the Cauchy criterion \eqref{eq:Cauchy-criterion}, whether there is a pole of $(\sigma - \chi(\sigma))^{-1}$ inside the triangle ABC. 
The code generating the results in this paper was implemented in C language with the ``complex.h" library.

Although the method was built specifically to tackle the problem of monokinetic beams presented in this paper, 
it can a priori be used in an efficient way to find the complex roots of any function $F(\sigma)$ which is well-behaved in the integration domains and such that the poles of ${F^{-1}(\sigma)}$ have a non-vanishing residue.

\bibliographystyle{jpp}

\begin{thebibliography}{99}
\bibitem{note1}{In the electrostatic description, 
     where the interaction is described with the Poisson equation, 
     the electric field is not even a dynamic degree of freedom. 
     Though this is not the most complete picture of the plasma, 
     it does account for several fundamental features of plasmas, 
     including Landau damping and Debye shielding.}
\bibitem{note2}{Formally, one may wish $N_s$ to be integer. 
    However, the linear dynamics \eqref{eq:C-wave}-\eqref{eq:A-wave}-\eqref{eq:Z-wave} for Fourier coefficients 
    no longer refers to individual particles, and any real positive values for $N_s/N$, $N$ and  $\varepsilon N$ 
    are mathematically admissible. Rational or integer values are not remarkable in this context.}
\end{thebibliography}

\end{document}